\documentclass[manuscript]{aastex}

\usepackage{graphicx}
\usepackage{epstopdf}
\usepackage{color}

\usepackage{natbib}
\def\comm#1 {{\tt (COMMENT: #1) }}

\newcommand{\tdep}{t_{depl.}}
\newcommand{\mugas}{\mu_{gas}}
\newcommand{\msun}{M_{\odot}}
\newcommand{\mstar}{M_{\star}}

\newcommand{\as}{^{\prime\prime}}

\shorttitle{Gas Fraction and Depletion Time at z$\sim$3.2}
\shortauthors{Schinnerer et al.}

\begin{document}


\title{Gas Fraction and Depletion Time of Massive Star Forming Galaxies at z$\sim$3.2: No Change in Global Star Formation Process out to $z>3$}


\author{E. Schinnerer\altaffilmark{1}}
\altaffiltext{1}{Max Planck Institute for Astronomy, K\"onigstuhl 17, 69117 Heidelberg, Germany}

\author{B. Groves\altaffilmark{2}}
\altaffiltext{2}{Research School of Astronomy and Astrophysics, Australian National University, Canberra, ACT 2611, Australia}

\author{M.T. Sargent\altaffilmark{3}}
\altaffiltext{3}{~Astronomy Centre, Department of Physics and Astronomy, University of Sussex, Brighton, BN1 9QH, UK}
 
\author{A. Karim\altaffilmark{4}}
\altaffiltext{4}{Argelander-Institut f\"ur Astronomie, Universit\"at Bonn, Auf dem H\"ugel 71, D-53121 Bonn, Germany}

\author{P.A. Oesch\altaffilmark{5,6}}
\altaffiltext{5}{Yale Center for Astronomy and Astrophysics, Department of Physics and Astronomy, Yale University, New Haven, CT 06511, USA}
\altaffiltext{6}{Observatoire de Geneve, Geneva University, 51 Ch. des Maillettes, 1290 Versoix, Switzerland}

\author{B. Magnelli\altaffilmark{4}}

\author{O. LeFevre\altaffilmark{7}}
\altaffiltext{7}{Aix-Marseille Universit\'{e}, CNRS, LAM (Laboratoire dÕAstrophysique de Marseille) UMR 7326, 13388 Marseille, France}

\author{L. Tasca\altaffilmark{7}}

\author{F. Civano\altaffilmark{8,9}}
\altaffiltext{8}{Yale Center for Astronomy and Astrophysics, 260 Whitney Avenue, New Haven, CT 06520, USA}
\altaffiltext{9}{Harvard-Smithsonian Center for Astrophysics, 60 Garden Street, Cambridge, MA 02138, USA}

\author{P. Cassata\altaffilmark{10}}
\altaffiltext{10}{Instituto de Fisica y Astronom'a, Facultad de Ciencias, Universidad de Valparaiso, 1111 Gran Bretana, Playa Ancha Valpara'so,
Chile}


\author{V. Smol\v{c}i\'{c}\altaffilmark{11}}
\altaffiltext{11}{Department of Physics, University of Zagreb, Bijeni\v{c}ka cesta 32, HR-10000 Zagreb, Croatia}


\begin{abstract}
The observed evolution of the gas fraction and its associated depletion time in main sequence (MS) galaxies
provides insights on how star formation proceeds over cosmic time. We report ALMA detections of the 
rest-frame $\sim$300$\mu$m continuum observed at 240 GHz for 45 massive ($\rm \langle log(M_{\star}(M_{\odot}))\rangle=10.7$),
normal star forming ($\rm \langle log(sSFR(yr^{-1}))\rangle=-8.6$), i.e. MS, 
galaxies at $\rm z\approx3.2$ in the COSMOS field. From an empirical calibration between
cold neutral, i.e. molecular and atomic, gas mass $\rm M_{gas}$ and monochromatic (rest-frame) infrared luminosity, the gas
mass for this sample is derived. Combined with stellar mass $\rm M_{\star}$ and star formation rate (SFR) estimates (from {\sc MagPhys}
fits) we obtain a median gas fraction of $\rm \mu_{gas}=M_{gas}/M_{\star}=1.65_{-0.19}^{+0.18}$ 
and a median gas depletion time $\rm \tdep(Gyr)=M_{gas}/SFR=0.68_{-0.08}^{+0.07}$; correction for the location on the MS will only
slightly change the values. The reported uncertainties are the 
$\rm 1\sigma$ error on the median.
Our results are fully consistent with the expected flattening of the redshift evolution from the 2-SFM (2 star formation mode) framework that 
empirically prescribes the evolution assuming a universal, log-linear relation between SFR and gas mass coupled to the redshift 
evolution of the specific star formation rate (sSFR) of main sequence galaxies. While $\rm \tdep$ shows only a mild dependence on 
location within the MS, a clear trend of increasing $\rm \mu_{gas}$ across the MS
is observed (as known from previous studies). Further we comment on trends within the MS and (in)consistencies with other studies.
\end{abstract}




\section{Introduction}

Knowledge of the gas fraction and depletion time of galaxies beyond the peak epoch of cosmic star formation rate (SFR) density, i.e. at redshifts $\rm z > 2$
is critical to determine their main mode of star formation and the efficiency of the star formation process. The study of galaxies 
in this redshift range is especially interesting in the context of potential discrepancies between 
the relative shape of the dark matter halo gas accretion rate and the measured evolution of the specific star formation 
rate (sSFR). For example, comparison between observations and predictions from cosmological simulations can test how efficiently 
accreted gas is incorporated into the gaseous disks
and converted to stars, e.g. gas fractions are fairly sensitive to the prescription used for the conversion of cold gas into stars 
\citep[see recent review by ][]{somerville15}. 

The evolution of the gas fraction and depletion time beyond z$\sim$2.5 is currently less well constrained than at lower redshifts
\citep[e.g. the recent compilation by][]{genzel15}, as much less objects have their cold gas fraction measured. Studies so far have focused
on small samples consisting of lensed galaxies 
\citep[e.g.][]{saintonge13,dessauges15} and non-lensed objects \citep{magdis12,tan14} for direct
detections. 
Normal star forming galaxies are observed to form a tight relation in the SFR vs. stellar mass plane that is often referred to as 
'main sequence' (MS) of star forming galaxies \citep{noeske07} whose normalization is a strong function of redshift \citep[for recent determinations at our redshift range of interest, see e.g.][]{tasca15}.
Recently, \cite{bethermin15} extended the analysis to $\rm z=4$ using infrared stacking and \cite{scoville16} presented the first direct
measurements for galaxies on and off the main sequence out to $\rm z\approx4.4$. 
These studies provide first important constraints on the 
gas fraction and depletion time suggesting that star formation in main sequence galaxies, i.e. normal star forming galaxies, proceeds in
a similar fashion to low redshift galaxies.

Probing the cold or molecular gas mass directly for $\rm z>3$ main sequence galaxies is very challenging even in the era of ALMA, as 
the expected strength of the CO emission lines requires long integration times. Furthermore, ALMA can typically only access high-J transitions
(CO(3-2) is the lowest transition accessible for such sources)
which are more prone to excitation effects and their relation to the bulk amount of molecular gas present is less straight-forward \citep[see, e.g.,][]{carilli13}.
The only possibility to quickly assemble sizable  samples is
to use the large bandwidth available for continuum detections and rely on
the observed tight relation between cold dust mass and neutral (i.e. molecular and atomic) gas mass \citep[e.g.][]{hildebrand83}.
This gas-to-dust ratio technique can either use the full information from the infrared spectral energy distribution or rely on a direct, locally 
calibrated relation between the sub-mm dust continuum and the cold gas mass \citep[e.g.][]{magnelli14,scoville14,groves15}. 
In general this method seems to lead to results that are in good agreement with CO-based gas masses \citep{genzel15} and has become
increasingly popular \citep[][]{magdis12,santini14,magnelli14,scoville14,scoville16,bethermin15}.
Both methods (via CO line or dust continuum) rely on local calibrations and thus exhibit a dependency on metallicity. Therefore studies
of high-redshift galaxies, particular on the low mass end, can be significantly affected by our ability to measure or statistically infer
gas phase metallicities \citep[e.g.][]{bethermin15}.

The sample, its properties and the data used for their determination are described in \S\ref{sec:sample}. The gas mass estimation and
results on the gas fraction and depletion time are presented in \S\ref{sec:results} and discussed in \S\ref{sec:discussion}. We summarize 
and conclude in \S\ref{sec:summary}.
Throughout the paper we assume a cosmology with $\rm H_0 = 70, \Omega_M=0.3$ and $\rm \Omega_{\Lambda}=0.7$
(for ease of comparison to other work in the literature) and use a Chabrier initial mass function for the stellar mass and SFR determination.


\section{Sample and Data}
\label{sec:sample}

\subsection{Sample Selection}

To study the cold gas fraction and the gas depletion time at a redshift of $z \sim 3-4$, we selected a 
sample of massive star forming galaxies in the COSMOS field to be observed with ALMA.
Our initial sample selection is based on the phot-z catalog of \citet{ilbert13} selecting all sources with good phot-z 
estimates ($\rm \Delta z < \pm0.2$) in the range $z_{phot} = 3.0 - 4.0$,  and masses determined by {\sc LePhare} 
\citep{arnouts99,arnouts02,ilbert06} with $\rm log(M_{\star}[M_{\odot}]) > 10.5$ (corresponding to UltraVISTA
magnitudes of $\rm K_s \lesssim 23\,mag$ \citep{mccracken12}). To 
select star-forming galaxies, 
we cross-matched our sample with the \citet{muzzin13} catalog and selected all objects with a MIPS 24$\mu$m counterpart 
of $\sim 3\sigma$ in the \citet{muzzin13} catalog, 
and with standard rest-frame UV-optical colors expected for star forming systems 
(based on the $J-[4.5]$(IRAC2) color versus M$_z$). This 
selection resulted in 73 sources. We added an additional 13 objects with spectroscopic
redshifts in our redshift and stellar mass range from a preliminary analysis of the VIMOS Ultra
Deep Survey \citep[VUDS, ][]{lefevre15} that matched our star-forming
requirements, resulting in a total of 86 potential target galaxies. 

For the analysis presented here we updated the photometry 
and redshift determination of this initial sample
based on the COSMOS2015 catalog of \citet{laigle16}. This catalog includes new $YJHK_s$ imaging from the UltraVISTA DR2 release, $Y$ band
imaging from Hyper Suprime-Cam and deeper SPLASH data at 3.6 and 4.5$\mu$m 
\citep[SPitzer Large Area  Survey, PI: Capak; ][]{steinhardt14}. 
The photometric redshifts are determined using the template-fitting {\sc LePhare} code 
based on the updated photometry using the fluxes computed within 3\arcsec\ apertures.
(Furthermore, we remeasured the MIPS$24\mu$m and PACS$100\mu$m fluxes; see \S \ref{subsec:sample-prop} for details.)

For 22 sources of our initial
sample, spectroscopic redshifts with high quality (flag $\rm \ge 2, >75$\% probability of being 
correct) exist to-date, 15 from VUDS \citep{lefevre15} and 7 sources mainly from zCOSMOS-deep \citep{lilly07}.  For 
these sources we use the spectroscopic redshifts instead of the photometric ones, which are consistent in all but 2 cases.
With these new redshifts, 7 of our original 86 sources have $z<2.8$. Excluding 
these sources, the median redshift of our sample is $z=3.2$  with the highest redshift being $z=3.8$.

\subsection{ALMA Data}

The 86 target fields were observed as part of an ALMA Cycle\,2 program (2013.1.00151.S, PI Schinnerer). The observations were 
optimized for continuum detections at 240 GHz (corresponding to $\sim$300$\mu$m rest-frame at the median sample redshift of 
$\rm z=3.2$) using the correlator 
in TDM mode with the four spectral windows centered at 231\,GHz, 233\,GHz, 247\,GHz, and 249\,GHz yielding a total bandwidth 
of 7.5\,GHz. The target fields were observed with typically 38 antennas between 25 and 30 of December 2014 for a total of 2.0min 
on-source integration time, with 24 fields receiving an additional 0.5min on-source time. The quasar J1010-0200 served as phase 
calibrator in all observations, for bandpass and flux calibration
J1058+0133, J0750+1231, J0854+201, J0825+0309, J1037-295, Ganymede, and Callisto were observed.
We used the calibrated data products provided by the ALMA project.
The resolution and rms achieved is 1.8''$\times$1.1'' (1.7''$\times$1.1'') and 66 (71) $\mu$Jy/beam for the 24 
(62) fields with (without) additional observing time when using natural weighting.
We CLEANed all four spectral windows together using the 'mfs' mode down to 3$\sigma$ without setting a CLEAN box and using up 
to 1,000 iterations. The final images cover an area of 39''$\times$39'' with a pixel size of 0.18'', sufficient to encompass the
primary beam FWHM of 23'' at 240 GHz.

\subsection{Stellar Mass and Star Formation Rate (SFR) Estimation}
\label{subsec:sample-prop}

Due to the up-dated redshift information for about 25\% of our objects and the better photometry available, we re-determined the stellar 
mass and SFR for each ALMA target. Given the expected large amounts of gas and dust present in our targets, we cannot rely on 
the UV emission alone to estimate the SFR due to the large, uncertain correction factors required. 
For all objects 24$\mu$m and 100$\mu$m fluxes or upper limits were extracted from the Spitzer/MIPS 24$\mu$m map 
from \citet{lefloch09} and the Herschel/PACS 100$\mu$m image
from \citet{lutz11} using the PSF fitting method of \citet{magnelli13} 
to simultaneously search for emission associated with the position of our targets, while also accounting for the flux from all known
24$\mu$m sources around it.
This careful de-blending is in particular necessary for sources \#26250, \#26318 and \#26388 as well as 
\#226676 and \#226748 which lie close together (within $\sim 10\as$, but still apart at 24/100$\mu$m resolution). 
This resulted in a significant fraction ($\rm \sim 30\%$) of targets without 3$\sigma$ MIPS$24\mu$m detections irrespective of the
source being ALMA detected or not. Only six sources are detected in the PACS$100\mu$m image above 3$\sigma$, half of these are 
detected by ALMA as well. This low detection rate is consistent with the depth of the PACS$100\mu$m image and the anticipated SFR of 
our targets.

Stellar masses and star-formation rates for the full sample are determined by fitting the available photometry within 
a 3'' aperture \citep[from the catalog of ][]{laigle16}, including our measured IR+ALMA fluxes and assuming the best available 
redshifts, with the {\sc MagPhys} 
code \citep{dacunha08}\footnote{In particular, we use the \citet{bruzual03} stellar 
libraries and the latest "high-z" version of the code available at \url{http://www.iap.fr/magphys/magphys/
MAGPHYS.html}.}. 
Non-detections by ALMA (see \S \ref{subsec:gas_mass}) are treated as upper limits. All PACS$100\mu$m photometry is also used
as upper limits due to their low S/N ($< 5\sigma$ in case of the detections) and potential confusion due to their low resolution. The MIPS$24\mu$m data photometry was excluded from the fitting due to the highly uncertain SEDs for star forming galaxies and potential contribution from an AGN.

Stellar masses determined with {\sc MagPhys} are consistent (barring a systematic 0.2\,dex offset) with {\sc LePhare}-based estimates in \citet{laigle16} 
for 90\% of the sample, and are independent (within uncertainty) of the inclusion of the IR+ALMA data.
Both UV and IR fluxes are used for our SFR measurements. When IR+ALMA photometry is excluded, SFRs are lowered by 0-0.5\,dex, indicating 
that significantly obscured star forming regions play a role in several of our sources. The use of the Bayesian SED fitting code {\sc 
MagPhys} allows us to include both these buried populations as well as the uncertainty of the IR SED which is mainly constrained by the 
PACS$100\mu$m upper limits and ALMA data points. We remind the reader that the optical/NIR data also provide a constraint though the rest-frame UV-optical shape to indicate possible extinction as indicated by the small offset mentioned above between the fits with and without the inclusion of the 
IR+ALMA data (though extremely heavily obscured young stellar populations are only constrained by the IR SED). 
We verified that our average SFR is
consistent with the mean and median SFR determined from the stacked IR SED (see \S \ref{subsec:gas_mass}). Moreover, the resulting average SED obtained via {\sc MagPhys} agrees very well with the stacked
SED. 
While the SFRs of individual sources will
suffer from the usual (systematic) uncertainties that are unavoidable for sources with poor sampling of their rest-frame infrared to sub-mm SED, the average SFR of our sample is hence robustly determined.

We have also identified 12 possible AGN in our sample by cross-matching with the Chandra COSMOS Legacy survey 
\citep{civano16}, and through a mid-IR excess \citep[i.e. red {[4.5]-[5.8] and [5.8]-[8.0]} colors, e.g. following the methodology of][]{lacy07}. 
Mid-IR excess AGN were identified via offsets in the IRAC photometry (in particular at  $8\mu$m) compared to  the fitted {\sc MagPhys} SED, 
which could not be reproduced by any reasonable stellar SED.
Two sources have clear (e.g. $\rm S/N > 5$) X-ray detections, 
while six more have weak X-ray detections, which we also classify as AGN. Seven sources have noticeable IR excesses, three 
of which also have X-ray detections.

\begin{figure}
\epsscale{.80}
\plotone{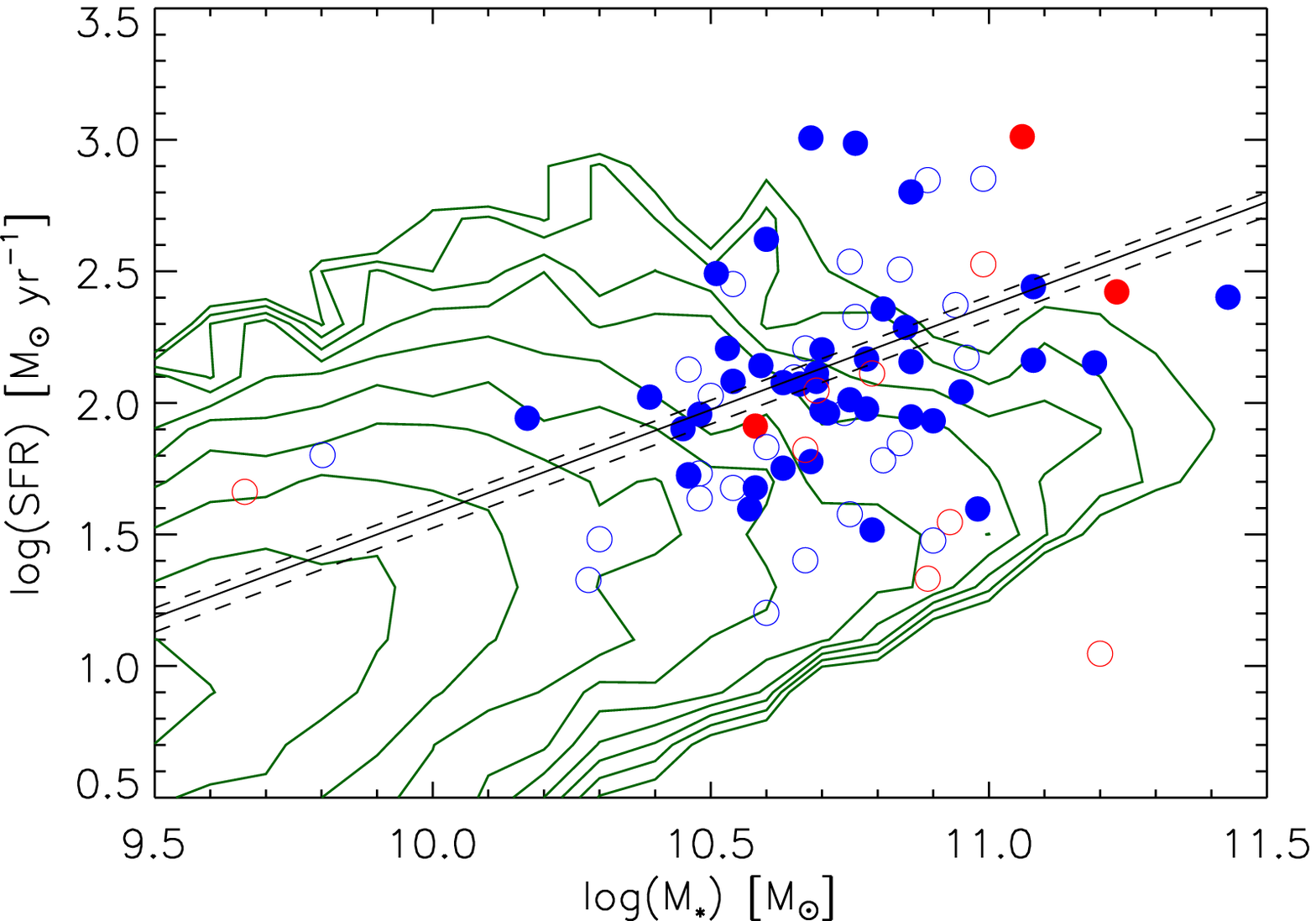}
\caption{Location of our $\rm z>2.8$ ALMA targets (detections: filled circles; non-detections: open circles) in the SFR vs. stellar 
mass $\rm M_{\star}$ plane compared to the distribution of all galaxies with good photometric redshifts at $\rm z\approx2.8-4.0$ 
in the COSMOS field (contours; $M_{\star}$ and SFR (both based on {\sc LePhare}) are taken from \citet{laigle16}). The solid line 
marks the location
of the main sequence of star forming galaxies at z=3.2 (dashed lines correspond to z=2.8 and 3.6, respectively) 
based on \citet{sargent14}. AGN are highlighted in red. (See text for details.)
\label{fig:sample}}
\end{figure}

In Fig. \ref{fig:sample} we show the location of our z$\sim$3.2 sample 
in the SFR vs. $\rm M_{\star}$ plane with respect to the location of the 'main sequence' of star forming galaxies at redshift $z\approx3$  
\citep{magdis10,lee11a,heinis14,bethermin14,tasca15,schreiber15,tomczak16}\footnote{In the compilation of main sequences
used by the 2-SFM framework \citep{sargent14} and up-dated with recent sSFR measurements \citep{heinis14,schreiber15,steinhardt14,bethermin14,tasca15},
galaxies with stellar masses of $\rm log(\mstar[\msun])\approx10.7$ (the typical mass of our
sample galaxies) show no evidence for an evolution of their sSFR values for $\rm 3 \le z \le 4$ and have an average sSFR of 
$\rm \sim 3\,Gyr^{-1}$ with an rms dispersion of 0.13\,dex. This is in good general agreement with other sSFR compilations 
\citep[e.g. by][]{speagle14}}.
Our ALMA detections sample the main sequence at high stellar mass ($\rm log(M_{\star}[M_{\odot}])\approx10.3-11.5$) and are not biased towards a particular SFR range ($\rm log(SFR [M_{\odot}yr^{-1}])\approx1.5-3$).


\section{Results}
\label{sec:results}

\subsection{240\,GHz Continuum Detection and Gas Mass Measurements}
\label{subsec:gas_mass}

For the detection of sources in our 86 target fields we use the source extraction software of \citet{karim13} that was developed 
for continuum source extraction in ALMA imaging data. The software automatically identifies sources at $\rm \ge 2.5\sigma$
and determines the flux based on a comparison of a three (assuming an unresolved source) and six parameter (assuming a 
resolved source) Gauss fit in the image plane. A total of 47 of our 86 targeted sources are 
detected at or above 3$\sigma$ (for the peak), with 45 (52\%) lying at a redshift of $\rm z>2.8$. We use the integrated fluxes from the 6 parameter fit and verified that it gives consistent results for unresolved sources.
We further tested that Gaussian fits at the position of the optical/near-IR sources result in similar values. For the ease of comparison
of serendipitous detections in our target fields in future work we use the values provided by the blind extraction software.
 
Based on the completeness tests done by \citet{karim13} up to 5\% of our detected sources (corresponding to $\sim$2 sources) 
could be spurious 
for a full sample of blind detections. Given that we are searching at predetermined positions, the probability is even lower, as our fields 
contain about 350 independent beams, so the chance that a spurious source would end up in the center is about 0.3\% (if there
is one source per field). The typical number of sources detected per field is about 7,
thus there is a $\sim$2\% chance that such a source would end up at the position of our targeted objects. Therefore we are
confident that all our detections are real. 

To convert the observed 240\,GHz flux density into a cold gas mass, we make use of the observed relation between cold dust luminosity 
and gas mass \citep[similar to, e.g.,][]{scoville14,scoville16}. Recently, \citet{groves15} calibrated relations between mono-chromatic IR luminosities
at 250, 350 and 500\,$\mu$m and neutral (atomic plus molecular) gas mass using high quality observations for 36 local galaxies
from the KINGFISH survey \citep{kennicutt11}. These 
calibrations implicitly include the effects of metallicity through the variation within the calibration sample. 
The calibration between monochromatic luminosity and gas mass becomes steeper and has larger scatter at shorter IR 
wavelengths because of the effect of dust temperature, and its correlation with stellar mass and metallicity. 
Here we adopt the relations for the high-mass sample, i.e. $\rm log(M_{\star}) > 9.0$ at rest-frame 250 and 350\,$\mu$m \citep[see Tab. 6][]{groves15} and
linearly interpolate the coefficients to match the observed rest-frame wavelengths:

\begin{equation}
\rm log_{10} (M_{\rm gas}(M_{\odot})) = (1.57 - 8\times10^{-4}\Delta\lambda) + (0.86+ 6\times10^{-4}\Delta\lambda)\times log_{10}(\nu L_{nu}(L_{\odot}))
\end{equation}

where

\begin{equation}
\rm \Delta\lambda = \lambda_{\rm ALMA,restframe} - 250\,\mu m
\end{equation}

and

\begin{equation}
\rm \nu L_{\nu}=\nu_{obs}\times S_{\nu,obs}\times4\pi\times D_{L}^2  .
\end{equation}

As discussed by \citet{groves15} the mono-chromatic IR luminosity relations
are similar to other methods advertised such as, e.g., the calibration of the 850\,$\rm \mu m$ luminosity by \citet{scoville14,scoville16}. We verified that 
gas masses derived using the \citet{scoville14} prescription are consistent with our results within the (systematic) uncertainties. 
The metallicities of high redshift galaxies are highly uncertain given the changes in the physical properties in these galaxies 
\citep[see e.g.][]{kewley13,kewley15}. Thus we assume that the metallicities of massive galaxies at our epoch of interest
 are broadly consistent with local objects of similar masses. However, there are suggestions that the metallicities of massive galaxies have increased since z$\sim$3 to now, with suggestions of around a factor of 2 \citep[e.g.][]{maiolino08,troncoso14,onodera16}. If this is the case our determinations for the gas mass of our sample are likely underestimated (by approximately 0.3\,dex), as we overestimate the metallicity and hence the Dust-to-Gas ratio in our detected sample. Note that the presumably higher dust temperature for galaxies at
 our redshifts \citep[e.g.][]{genzel15,bethermin15} compared to the local sample could counteract such a systematic trend.

As a further confirmation, we also derived the average gas mass of our detected sources in the redshift range z=2.8 to 3.7 through determining the gas mass 
from stacking of the (Spitzer and Herschel) infrared data centered on our detected source positions. 
We use the methodology of \citet{magnelli14} 
as used for \citet{genzel15}, i.e. a modified \citet{draine07} model is fitted to the IR 
data points plus the average 240\,GHz ALMA flux density, providing a mean dust temperature and dust mass for the sample. 
The average dust mass obtained with this approach is then converted into a gas mass by applying the metallicity dependent 
dust-to-gas ratio for $z\sim0$ star-forming galaxies found by \citet{leroy11}.
The average metallicity of our detected sources is derived using the stellar mass-metallicity relation at their average redshift as 
found in \citet{genzel15}. 
Comparison of the average metallicity of our detection sample based on the \citet{genzel15} prescription \citep[which uses the mass-metallicity relation of][]{maiolino08} yields very similar results (within $<0.1$\,dex) as more recent determinations of the mass-metallicity relation \citep[e.g.][]{troncoso14,onodera16}.

The mean gas mass derived from the fit to the stacked infrared SED of $\rm \langle M_{gas}(\msun)\rangle_{IR SED}=10.89$ 
(with a full accounting of the potential effect of the mass-metallicity relation)
is remarkably similar to the mean gas mass from the mono-chromatic approach of $\rm \langle M_{gas}(\msun)\rangle_{L300}=10.84$ .
We take this as an indication that no strong systematic biases are present between results from these two approaches when 
comparing the same high-mass objects.

For the subsequent analysis
we restrict the sample to objects with ALMA detections in the redshift range of z=2.8 to 3.7.
These objects have a robust determination of their SFRs due to the inclusion of the ALMA fluxes, as our re-measuring
of the IR photometry for all objects sometimes led to a significant change in the MIPS$24\mu$m flux.
These 45 objects sample fairly well the 
SFR versus stellar mass $\rm M_{\star}$ plane (see also Fig. \ref{fig:sample}), and have a mean redshift of z=3.2, a mean stellar mass of 
$\rm \langle log(M_{\star}[M_{\odot}])\rangle=10.7$ and a mean SFR of $\rm \langle log(SFR[M_{\odot}yr^{-1}])\rangle=2.1$. The 
average specific SFR (sSFR) of our sample is $\rm log(sSFR[yr^{-1}])= -8.6$, so our sample lies close to the main sequence at the mean 
redshift of our sample (with a median offset of -0.04\,dex, i.e. very slightly below the MS, but well within the MS scatter).

\subsection{Gas Fraction and Gas Depletion Time}
\label{subsec:gas}

In the following we combine the estimates of stellar mass, SFR and gas mass for our 45 detections at $\rm 2.8 \le z < 4$ 
to study the evolution of the 
gas fraction, defined as $\rm \mu_{gas} = M_{gas}/M_{\star}$
and gas depletion time, defined as $\rm \tdep = M_{gas}/SFR$. 
We compare our measurements to lower redshift results available from the literature as well as empirical
predictions for main sequence galaxies (at our target redshift) from the 2-SFM model \citep{sargent14}.  
The 2-SFM predictions are based on (a) a log-linear, redshift-independent star-formation law (calibrated for $\rm z<2.5$ MS galaxies) 
relating SFR and molecular gas mass, and (b) the observed redshift-evolution of the specific SFR of main sequence galaxies.
We remind the reader that values derived for individual galaxies have significant uncertainties and might suffer from
systematic uncertainties. However, given the independent cross-checks on the average properties done in \S \ref{subsec:sample-prop} and \ref{subsec:gas_mass},
we expect that average trends described in the following are robust.

\subsubsection{Relevance of Being on the main sequence of Star Forming Galaxies}
\label{subsec:sSFR}

\begin{figure}
\epsscale{1.0}
\plotone{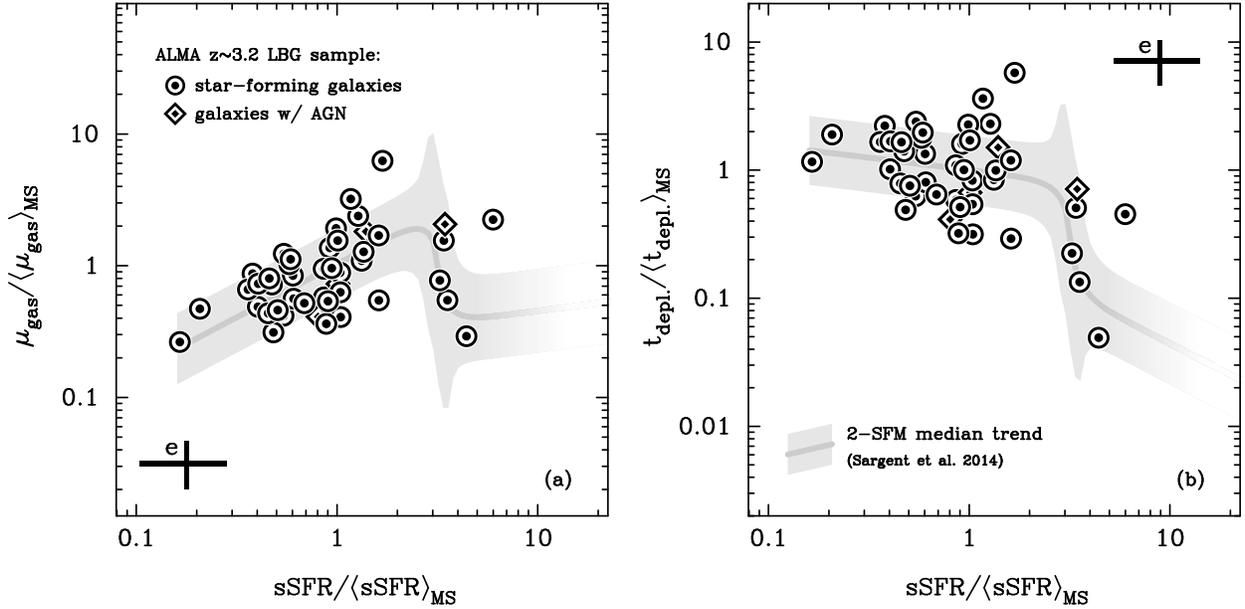}
\caption{Gas fraction and depletion time versus specific SFR (sSFR) of our z=3.2 sample relative to average main sequence properties. 
{\it Left, panel (a):} The gas fraction $\rm \mu_{gas}=M_{gas}/M_{\star}$ relative to the average gas fraction of main sequence 
galaxies $\rm \langle\mu_{gas}\rangle_{MS}$ of our sample is shown as black symbols. 
{\it Right, panel (b):} The gas depletion time $\rm \tdep=M_{gas}/SFR$ relative to the average gas depletion time of
main sequence galaxies $\rm \langle\tdep\rangle_{MS}$ of our sample is presented by black symbols. Circles and diamonds indicate
star forming galaxies and candidate AGN hosts. The cross marked with 'e' represents the typical error bars in each panel.
It includes the scatter of sSFR measurements as reported in the literature, i.e. it also illustrates the systematic uncertainties relating to 
the exact normalization of the MS and also the best-fit Schmidt-Kennicutt relation. 
The dark gray line shows the predicted, redshift-invariant median trends from the 2-SFM framework \cite{sargent14} 
as a function of distance from the main sequence of star forming galaxies (with the gray band spanning the expected 1$\sigma$ 
scatter around the average). The grey coloring fades out toward the extreme starburst regime which is not relevant to this study.
Note that for low-z galaxies the contribution of atomic (HI) gas is not included in $\rm M_{gas}$.
\label{fig:sSFR}}
\end{figure}

In Fig.\,\ref{fig:sSFR}\,(a) we show the deviation of the measured gas fraction $\rm \mugas$ relative to the 
expected\footnote{Given the stellar mass of an 
average MS galaxy its SFR is determined by the sSFR-evolution of MS galaxies. The (molecular) gas mass of this
galaxy is given by the trend line of the integrated $\rm M_{\star}-M_{gas}$ relation in Fig. \ref{fig:KSlaw}. The stellar
and gas masses have then been combined to form the expected gas fraction$\rm \langle\mu_{gas}\rangle$ if a `typical' MS galaxy.} gas  
fraction ($\rm \mu_{gas}/\langle\mu_{gas}\rangle_{MS}$) of a star forming galaxy located exactly on the mean main sequence locus.
Gas fractions are predicted (grey shaded areas in Fig.\,\ref{fig:sSFR}) to vary significantly as a function of offset $\rm sSFR/<sSFR>_{MS}$
from the main sequence. In keeping with expectations, we do see a trend in our data that galaxies below the main sequence, i.e.
with lower SFR for a given stellar mass, 
exhibit a lower gas fraction while galaxies above the main sequence have a higher gas fraction. Over the range of 
specific SFR probed by our galaxies ($\rm \pm0.7\,dex$ or from 20\% to 500\% of the main sequence value), their gas fractions range 
over 1.5\,dex (from $\sim20\%$ to 600\% of the average main sequence value). This trend is consistent with the 2-SFM predictions,
thus implying a close correlation between SFR and $\rm M_{gas}$ as observed in $\rm z<2.5$ galaxies.  

A similar plot is shown for the deviation of the depletion time relative to the average depletion of the main sequence 
$\rm \tdep/\langle\tdep\rangle_{MS}$ in Fig.\,\ref{fig:sSFR}\,(b). As expected from the correlation seen in 
Fig.\,\ref{fig:sSFR}\,(a), the variation as a function of distance from the main sequence is less pronounced for sources with values
less than three times the MS value.
Again, the data mainly follow the predicted trend from the 2-SFM model, also for sources in the transition to the starburst regime.
The scatter for $\rm \tdep$ and $\rm \mu_{mol.}$ is consistent with that seen at lower redshift used to predict the 2-SFM distribution. 
No obvious differences in the distribution of star forming galaxies and galaxies potentially hosting an AGN can be seen for either quantity.

A clear link between $\rm \mugas$ and sSFR has already been reported for local and high redshift (z$<3$) galaxies 
\citep[e.g.][]{saintonge12,magdis12,bothwell14,genzel15,dessauges15}. Our data suggests that the steep trend in $\rm \tdep$ with sSFR 
reported in the literature \citep[e.g.][]{saintonge11b,genzel15,dessauges15} could at least partially be due to the transition to more 
starburst-like objects
as indicated by the 2-SFM model predictions and the distribution of our few sources. As the transition region in the 2-SFM model encompasses about 1 orders of magnitude and small errors on the sSFR
determination could lead to large scatter,
dedicated high quality observations will be required to further test this.

\subsubsection{Time Evolution}

\begin{figure}
\epsscale{1.0}
\plotone{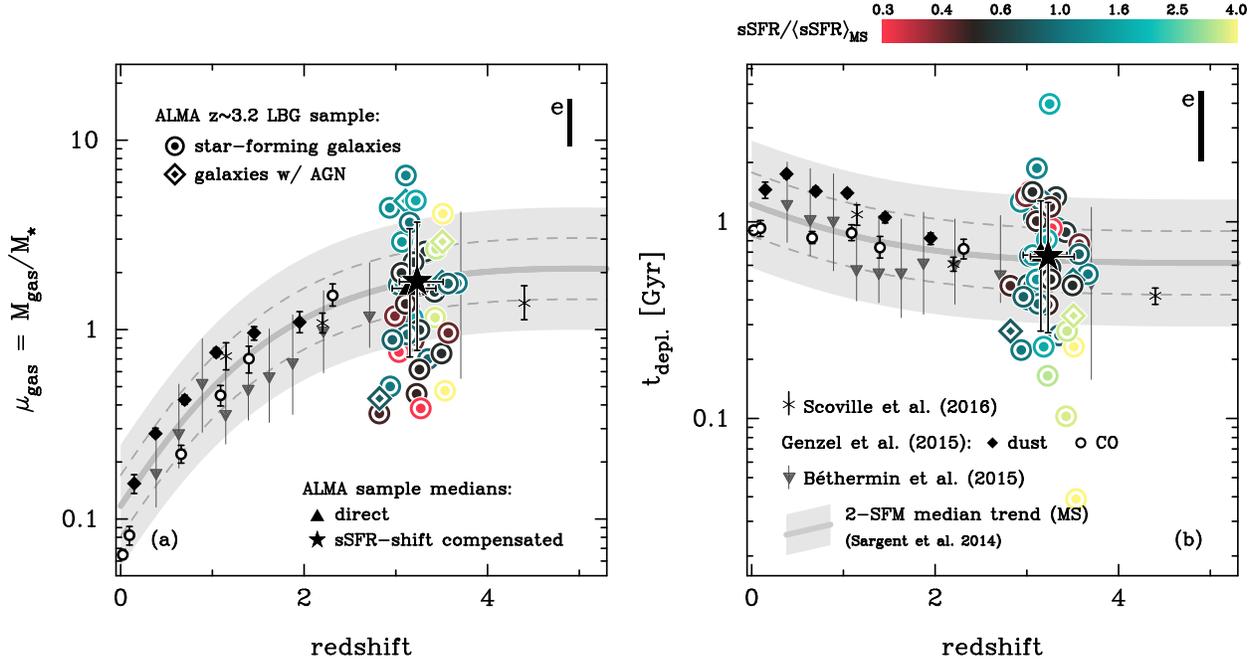}
\caption{Redshift evolution of the (a) gas fraction $\rm \mu_{gas}=M_{gas}/M_{\star}$ ({\it left}) and (b) depletion time $\rm t_{dep}$ 
({\it right}). 
Our sample is color-coded according to the distance from the main sequence of star forming galaxies with redder (greener) colors indicating 
objects below (above) the main sequence (see color bar in the top right) separated into star forming galaxies (circles) and
candidate AGN hosts (diamonds).  
The typical error bar of our objects marked by an 'e' is indicated in the top right
corner of each panel. The median values of our z=3.2 sample are shown by large black triangles, while the median
values compensated for the off-set of our sample from the main sequence are given by the large black star symbols.
The median trend predicted by the 2-SFM model for main sequence galaxies \citep{sargent14} is shown by the solid grey line, 
 the dashed lines indicate the expected gas fraction (depletion time) for galaxies lying 1$\sigma$ above or below the 
 star-forming main sequence. Galaxies within 2$\sigma$ of the average main sequence locus are predicted to lie 
 within the light grey band.
Note that for low-z galaxies the contribution of atomic (HI) gas is not included in $\rm M_{gas}$.
In addition, the average gas fraction derived from CO line observations (black open circles) and from stacked 
dust SEDs (black diamonds) at lower redshifts from the compilation by \citet{genzel15} is shown together with
recent results using stacked dust SEDs in the COSMOS field \citep[dark grey triangles, ][]{bethermin15} and 
sub-mm continuum measurements \citep[black filled circles, ][]{scoville16}.
\label{fig:redshift}}
\end{figure}

We compare the gas fraction $\rm \mu_{gas}=M_{gas}/M_{\star}$ for our $\rm z\sim3.2$ sample
to the predicted evolutionary trends from the 2-SFM model as well as literature values for larger samples 
\citep{genzel15,bethermin15,scoville16} in Fig.\,\ref{fig:redshift}\,(a). Comparison
to literature values from large samples shows that our results are in very good agreement with the findings of \citet{genzel15}, and 
\citet{bethermin15}. Our galaxies are also consistent with the average MS values of \citet[][their Tab. 2]{scoville16}.
The median value of our z=3.2 sample of $\rm \langle\mugas\rangle=1.65_{-0.19}^{+0.18}$ 
(filled black triangle; uncertainties are 1$\sigma$ error on median) and the one compensated for the MS off-set of our 
sample of $\rm \langle\mugas\rangle_{comp}=1.68_{-0.19}^{+0.18}$ (filled black star; uncertainties are 1$\sigma$ error on median) 
are consistent with the expected flattening of the $\mu_{gas}(z)$ curve at $\rm z\gtrsim2.5$ in
the 2-SFM model \citep{sargent14}. The $\rm z=4.4$
data point of \citet{scoville16} falls slightly below the 2-SFM model line. As \citet{scoville16} also include
non-detections in their average values \citep[unlike][for the CO line measurements]{genzel15}, this could point to a certain biasing of the
results when only detections are considered \citep[however, see the stacking data points from][]{bethermin15} 
or be due to small number statistics \citep[][include only 6 galaxies in their $\rm z=4.4$ measurement]{scoville16}.

Our objects show a trend of gas fraction with specific SFR (as highlighted by the symbol color used in Fig. \ref{fig:redshift}
and expected from \S\,\ref{subsec:sSFR}) covering more than one order of 
magnitude at z$\approx$3.2. An exception are those galaxies that lie more than $2.5\times$ above the MS and do not follow the same 
trend due to their significantly different $\rm \tdep$ or star formation efficiency ($SFE=\tdep^{-1}$) compared to MS galaxies. Our three AGN host 
galaxy candidates follow the overall trend of the remaining 42 galaxies.

This demonstrates how sensitive gas fraction measurements are on the range of SFRs sampled and illustrates how varying selection 
criteria in different studies may produce discrepant results. Sources located below the main sequence of star
forming galaxies, i.e. with SFR lower than the median galaxy, will bias the gas fraction toward lower values.
It also underlines the importance of a good (or at least
consistent) determination of the SFR when comparing different surveys or in other words the strong relation between $\rm M_{gas}$ and SFR 
strongly affects the interpretation of $\rm \mu_{gas}$. 
Our objects closest to the main sequence (color-coded dark green to black) typically lie closest to the expected value of $\rm \mu_{gas}
\approx2$, 
but they still show a significant scatter suggesting that the intrinsic gas fraction for galaxies on the main sequence has a
wide distribution. (Note that some of the scatter is also due to the fact that our galaxies sample a range of stellar masses
with different average gas fractions; see Fig.\,\ref{fig:mstar}.) Larger samples are required to confirm this behavior.

Comparison of the gas depletion time $\rm \tdep$ of our $\rm z\sim3.2$ sample to the 2-SFM model trend \citep{sargent14} and literature 
values for larger samples is presented in Fig.\,\ref{fig:redshift}\,(b). 
The median direct and compensated values of our z=3.2 
sample of $\rm \langle\tdep(Gyr)\rangle=0.68_{-0.08}^{+0.07}$ and $\rm  \langle\tdep(Gyr)\rangle_{comp}=0.67_{-0.08}^{+0.07}$ 
(uncertainties are 1$\sigma$ error on median) are fully consistent with the trends and values of \citet{genzel15} and 
\citet{bethermin15} which also follow the trend 
predicted by the 2-SFM model. 
According to data and model, $\rm \tdep$ may approach a 
plateau of $\rm \tdep\approx600\,Myr$ for MS galaxies at $\rm z \ge 3$. 
Again, the mean values of the MS galaxies from \citet{scoville16} are fully consistent with the other data,
with the exception of the $\rm z=4.4$ data which tends to values lower than predicted by the 2-SFM model, but still consistent within
the expected scatter.

Our objects show no clear trend with specific SFR (as expected from \S \ref{subsec:sSFR}) over almost two orders of magnitude at 
z$\approx$3.2. An exception are the objects that lie more than $\rm 2.5\times$ above the main sequence that all exhibit depletion times
of 300\,Myr and less. It is interesting to note that our scatter is similar to that from the IR stacking analysis by
\citet{bethermin15}. This suggests that in addition to potential sample selection biases (fraction of starburst-like sources), 
a large intrinsic scatter could contribute to variations between different studies.

\subsubsection{Trends with Stellar Mass}

\begin{figure}
\epsscale{.80}
\plotone{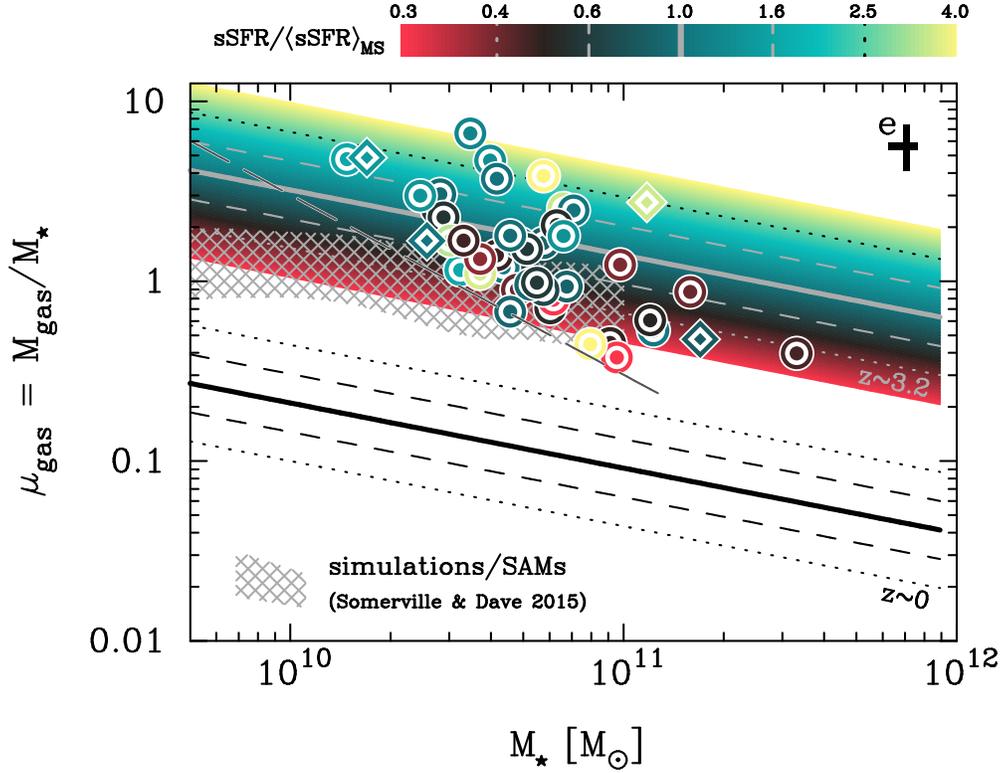}
\caption{Gas fraction $\rm \mugas$ versus stellar mass $\rm \mstar$ at $\rm z\sim3.2$. Our z=3.2 sample is color-coded
based on distance from the main sequence of star forming galaxies at $\rm z\sim3.2$ (see color bar on top of panel). 
Star forming galaxies are indicated as circles, potential AGN hosts as diamonds. The typical error bar marked with an 'e'
is shown in the top right corner. The
completeness limit (dark gray, dashed line) is dictated by our 3$\sigma$ flux limit at $\rm z\approx3.2$, $\rm log(M_{gas}[M_{\odot}])=10.48$. The predicted trend of gas fraction $\rm \mugas(\mstar)$ from 2-SFM
\citep{sargent14} is shown with the same color-coding as our data points, the solid grey line gives the median trend (for remaining lines see
color bar). 
The corresponding lines in black show the same information for local galaxies at $\rm z\sim0$. The range of possible values 
predicted by cosmological simulations and semi-analytical models (SAMs) for MS galaxies compiled by \citet{somerville15} is
shown as the grey cross-hatched area.
Note that for low-z galaxies the contribution of atomic (HI) gas is not included in $\rm M_{gas}$.
\label{fig:mstar}}
\end{figure}

\begin{figure}
\epsscale{.50}
\plotone{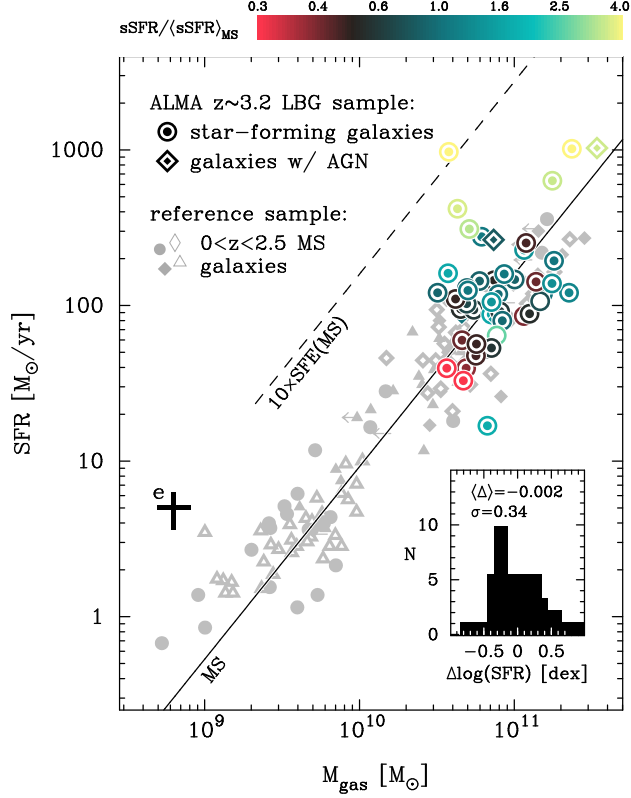}
\caption{Star formation rate (SFR) versus (molecular) gas mass. Our z=3.2 sample (point color-coded based on distance from the 
main sequence of star forming galaxies; color bar on top) scatters around the same power-law relation between SFR and gas mass 
occupied by $\rm 0<z<3$ main sequence galaxies (marked 'MS'). For reference the location of high-efficiency starburst galaxies 
with ten-fold enhanced SFE (star formation efficiency) compared to MS galaxies is shown as well (black line labelled
 '10$\times$SFE(MS)').
Star forming galaxies are shown as filled circles, while 
candidate AGN hosts as filled diamonds (open symbols are for $\rm z<2.8$ sources only) A representative error bar is indicated in the bottom left corner marked
by an 'e'. The reference sample is from the compilation of \citet{sargent14} and
comprises local to $\rm z\sim2.5$ galaxies on the main sequence. 
Note that for low-z galaxies the contribution of atomic (HI) gas is not included in $\rm M_{gas}$.
\label{fig:KSlaw}}
\end{figure}

The gas fraction $\rm \mugas$ is slightly anti-correlated with stellar mass $\rm \mstar$ for our z=3.2 sample as shown in Fig.\,\ref{fig:mstar}.
Our galaxies agree well with the general predicted trend by the 2-SFM model \citep{sargent14}, 
but they reveal an apparently steeper slope of $\rm \mugas(\mstar)$ that is likely an artifact created by our detection limit (see
long-dashed diagonal line in Fig.\,\ref{fig:mstar}). 

At fixed stellar mass, galaxies with high gas fractions are on average expected to correspond to MS galaxies well above the mean MS 
locus according to the 2-SFM framework (see colored band in background of Fig.\,\ref{fig:mstar}). In Fig.\,\ref{fig:mstar} we have colored our 
ALMA detections according to the actual (s)SFR-offset with respect to the MS. While there is an overall, broad agreement between 
expected and actual location of our data in Fig.\,\ref{fig:mstar} (in the sense that high-sSFR galaxies lie above the average $\rm \mugas$ vs. 
$\rm M_{\star}$ trend and low-sSFR galaxies below), the fact that the color of the individual symbols and the background color scheme do 
not match up perfectly evidences a non-neglible scatter of the SFR vs. $\rm M_{gas}$ relation. We will discuss this further in 
\S\,\ref{sec:discussion}. Note that galaxies with $\rm sSFR/\langle sSFR\rangle_{MS}\gtrsim3$ have gas fractions similar to 
the average MS galaxy with identical mass due to their enhanced star formation efficiency (or lower $\rm \tdep$; see also 
Fig.\,\ref{fig:sSFR}).
It is interesting to note that most of the $\rm log(M_{\star}(M_{\odot}))>11$ galaxies are below the
MS and also have the lower $\rm \mugas$ expected for objects with lower sSFR than the MS.
However, the overall scatter is consistent with the range expected from the MS range probed by our galaxies.

Comparison of the trend from the 2-SFM model and our data to predicted relations from cosmological simulations and 
semi-analytical models (SAMs) compiled by \citet{somerville15} shows that the observed gas fraction is typically a factor 
of $\sim 2\times$ higher. The dependence of $\mugas$ on $\mstar$ is probably stronger than seen in the cosmological
models, though results appear to be very sensitive to the exact MS location of the sample studied. Our AGN host galaxies
cover no preferred parameter space. 
A similar trend of $\rm \mugas$ is seen locally \citep{saintonge11a,bothwell14} and also at higher redshift \citep[e.g.][]{genzel15,dessauges15}. 
No trend is obvious in our data when plotting $\rm \tdep$ against $\mstar$. Since our data cover the same parameter space as the 
$\rm z>1$ star forming galaxies compiled by \citet[][their Fig. 9]{dessauges15} this might be an effect of our limited stellar mass range
probed.


\section{No change in global star formation process out to $z>3$}
\label{sec:discussion}

The results from section \S\ref{subsec:gas} for $\rm \tdep$ and $\rm \mugas$ are (i) a redshift independent dependance on sSFR and
(ii) a flattening of the redshift evolution at $z\sim3$.
They imply that our z$\sim$3 massive star forming galaxies follow the relation
between SFR and gas mass (i.e. the Schmidt-Kennicutt relation) established for lower redshift galaxies. This relation is one key ingredient in the 2-SFM 
prescription \citep[see ][]{sargent14}.
Placing all our detections onto the SFR vs. gas mass plane (Fig.\,\ref{fig:KSlaw}) shows that the z$\sim$3 MS galaxies
occupy the same space as lower redshift objects at $\rm 1.2<z<2.5$. 
The amplitude of the scatter of our sample restricted to $\rm \pm0.5\,dex$ from the MS location
(see inset in the bottom right corner) is about 1.7$\times$ larger than that observed for the reference MS samples (of $\approx$0.20\,dex).
It is interesting to note that there is a slight trend with MS location: galaxies
well below the MS have a higher depletion time $\rm \tdep$ (or lower star formation efficiency) 
than objects well above the MS that typically have a lower $\rm \tdep$ (or higher star formation efficiency)
than objects close to the MS. This is consistent with the trend seen in Fig.\,\ref{fig:sSFR}\,b. 
The larger scatter in SFR is mainly caused by galaxies in the transition to starburst objects.
The AGN host galaxies fall within the scatter of our star forming galaxies. The mean MS values of \citet{scoville16} are also
consistent with the location of our data points.

Our $z=3.2$ MS galaxies follow the Schmidt-Kennicutt relation as determined through the fit by \citet{sargent14}. This explicitly means that 
a single star formation prescription or 'law' describes the relation between molecular gas mass and star formation activity out to $z>3$ -- at least for MS galaxies. 
This was already concluded by \citet{bethermin15} from their IR stacking analysis. 

The proposed redshift evolution of \citet{saintonge13} for $\rm \tdep$ of $(1+z)^{\alpha}$ (their Eq. 10) of $\rm 0.36-0.17\,Gyr$ 
(for their $\rm \alpha\in[-1.0,-1.5]$) is inconsistent with our median $\rm \tdep$ of  0.67\,Gyr by a factor of $\gtrsim2\times$.
In the equilibrium framework developed for the bath tub or reservoir model, it is assumed $\rm \tdep=t_H\,M_{star}^{-0.3}$ \citep{dave12}.
As the literature and our data all probe roughly MS galaxies with similar stellar masses of a few times $\rm 10^{10}\,\msun$, the dependence
should simplify to $\rm \tdep\sim t_{H} \sim t_{dyn} \sim(1+z)^{-1.5}$ \citep[e.g.][]{dave12}. Based on the data available it seems that
this assumption is not valid or no longer valid at $z>3$.
For $\rm \mugas$ the situation is less clear as our preferred redshift-independent dependence of $\rm \mugas$ on sSFR
from the 2-SFM framework gives values consistent with $\rm \mu_{gas}=1/(1+(\tdep(z)*sSFR(z))^{-1}))$ \citep[see, e.g., Fig. 11 of][]
{saintonge13}. As this equation is
one of the equilibrium relations proposed by \citet{dave12} this implies that sSFR might evolve less strongly with redshift than assumed, i.e.
a shallower evolution than $sSFR \sim (1+z)^{2.25}$ expected from cosmic inflow driven by gravitational infall \citep{dave12}
or the assumption of $\rm t_{dyn} \sim (1+z)^{-1.5}$ is not correct implying that the relation between galaxy and halo mass would
evolve.
Similarly, the trends for $\rm \mugas$ ($\rm \tdep$) determined by \citet{genzel15} are not consistent at $\rm z\approx3.2$ with the measurements for our MS galaxies. 
In any case observations at redshifts higher than $\rm z=3$ are required for a more definite answer. 

An interesting consequence of the fairly constant $\rm \tdep$ and $\rm \mugas$ at $z>2$ is the implication that the evolution in
the cosmic star formation rate density at $\rm z>2$ for the high mass systems studied here is then driven by the number density of such star forming systems and not
a change of the gas reservoir available. This is in contrast to the explanation for the observed strong decline in the cosmic star formation
density at $z<1$ where \citet{karim11} argue that a decline in the cold gas reservoir is the root cause as the number density of (massive)
star forming systems does not change. This might imply that we see the transition from the epoch of galaxy formation (at high z) to the
epoch of star formation (at low z). The observed strong evolution of the UV luminosity function at $\rm z>4$ \citep[e.g.][]{bouwens15}
suggests that the drop in the cosmic SFRD at these redshifts might be due to a combination of a decreasing number density of star forming systems and an
evolution of their luminosity. If the luminosity evolution of the high mass systems is not as strong as inferred from the UV work, as suggested by \citet{rowan16} based on an analysis of IR data, our original statement made for objects in the range $\rm 1.5<z\lesssim3$ can be extended to higher z.

\citet{lagos15} show the expected evolution of $\rm \mugas$ and $\rm \tdep$ (their Fig. 11) for two different prescriptions of $\rm H_2$ 
formation applied to a set of the EAGLE cosmological simulations. Our observed values are about 3$\times$ higher for both parameters than
the predictions for MS galaxies with $\rm log(M_{\star}(M_{\odot}))>9.7$ based on the simulations. \citet{lagos15}
explain the drop in gas fraction and depletion time beyond z$\approx$3 with a lower $\rm H_2$ formation efficiency than at $z<1$ due to 
lower gas phase metallicity and higher star formation surface densities. The mismatch might be due to the fact that our observations are 
probing a higher mass range implying a strong dependence of $\rm H_2$ formation efficiency on stellar mass or that some physics are not 
sufficiently
captured by the simulations.

Our data allows for a first glimpse on a potential evolution within the main sequence. \citet{tacchella16} observe in their simulations the f
ollowing trend : Galaxies above the MS compactify their gas reservoir leading to higher SFR, higher $\rm \mugas$ and 
shorter $\rm \tdep$. Once the (central) gas reservoir is exhausted, galaxies drop below the MS with lower SFR, lower $\rm \mugas$ and 
longer 
$\rm \tdep$. Our data is consistent with a MS trend in $\rm \mugas$, however, a MS trend for $\rm \tdep$ is only evident once objects in 
transition to starbursts are taken into account. Thus to test this scenario further a larger sample and, more critically, morphological 
information on the gas reservoir, i.e. its size, are required.

In summary our ALMA observations of the gas mass in 45 z$\approx$3.2 massive MS galaxies suggest that a single relation between gas 
mass
and SFR is sufficient to explain the evolution of the gas fraction and depletion out to $z>3$ for normal star forming galaxies. As already 
pointed out earlier our methods accounts for the contribution of molecular and atomic gas, while most of the other methods are calibrated 
for the molecular component only. Given the close agreement between derived gas masses of our method and the calibration of 
\citet{scoville16} for our galaxies as well as the expected low atomic gas content at $z>3$ \citep[e.g.][]{obreschkow09,popping14,lagos11}, we do not expect 
that this introduces large uncertainties for our sample galaxies.


\section{Summary and Conclusions}
\label{sec:summary}

We present 45 ALMA continuum detections of massive main sequence (MS) galaxies (with $\rm \langle log(M_{\star}(\msun)\rangle=10.7$
and $\rm \langle log(sSFR(yr^{-1})\rangle=-8.6$)
at $z\approx3.2$ in the COSMOS field. 
Conversation of the rest-frame sub-mm continuum luminosity to gas mass allowed us to derive the gas fraction 
$\rm \mugas=M_{gas}/M_{\star}$ and gas depletion time $\rm \tdep=M_{gas}/SFR$. Our data points are consistent with literature
values reported at lower and higher redshifts for these parameters. Our median values of $\rm \langle\mugas\rangle=0.68$
and $\rm \langle\tdep\rangle=0.68\,Gyr$ imply a flattening of the redshift trends beyond $z\approx2$ inconsistent with the expected
evolution for a strong dependence on dynamical time. The surprisingly good agreement with the predicted trends from the 2-SFM framework
suggest that these analytic prescriptions are a good representation of the evolution of $\rm \mugas$ and $\rm \tdep$ -- at least for massive
MS galaxies out to $z\sim4$. It further implies that the relation between SFR and gas mass is constant and does not evolve over time.

\acknowledgments

We warmly thank the German ALMA ARC node for their excellent support for this project and R. Somerville for providing the information
on the simulation predictions for our redshift range of interest.
We also thank the referee for constructive comments.
ES acknowledges fruitful discussions with R. Somerville, R. Feldmann, N.Z. Scoville, N. Bouch\'{e}, and C. de Breuck that helped to significantly focussed the paper.
B.G. gratefully acknowledges the support of the Australian Research Council as the recipient of a Future Fellowship (FT140101202). 
A.K. acknowledges support by the Collaborative Research Council 956, sub-project A1, funded by the Deutsche Forschungsgemeinschaft (DFG)
Support for B.M. was provided by the DFG priority program 1573 ``The physics of the interstellar medium''.
VS acknowledges funding from the European Union's Seventh Frame-work program under grant agreement 337595 (ERC Starting Grant, 'CoSMass'). 
This paper makes use of the following ALMA data: ADS/JAO.ALMA\#2013.1.00151.S. ALMA is a partnership 
of ESO (representing its member states), NSF (USA) and NINS (Japan), together with NRC (Canada), NSC 
and ASIAA (Taiwan), and KASI (Republic of Korea), 
in cooperation with the Republic of Chile. The Joint ALMA Observatory is operated by ESO, AUI/NRAO and NAOJ.
Based on data obtained with the European Southern Observatory
Very Large Telescope, Paranal, Chile, under Large Program 185.A-0791.

{\it Facilities:} \facility{ALMA}

\clearpage

\bibliography{z_gt_3}


\clearpage


\clearpage


\clearpage


\clearpage


\clearpage

\end{document}